\documentclass{moriond}

\usepackage{latexsym}
\usepackage{epsfig}
\usepackage{graphicx,subfigure}
\usepackage[normalem]{ulem}
\usepackage{color}
\usepackage{xcolor}
\usepackage{multirow}
\usepackage{hyperref}
\usepackage{numprint}
\usepackage{eucal}
\usepackage{amsmath}
\usepackage{amssymb}
\usepackage{wasysym}

\newcommand*{\mathabxbfamily}{\fontencoding{U}\fontfamily{mathb}\selectfont}
\DeclareFontFamily{U}{mathb}{\hyphenchar\font45}
\DeclareFontShape{U}{mathb}{m}{n}{
      <5> <6> <7> <8> <9> <10> gen * mathb
      <10.95> mathb10 <12> <14.4> <17.28> <20.74> <24.88> mathb12
      }{}

\newcommand*{\Earth}{{\text{\mathabxbfamily\char"43}}}

\def\be{\begin{equation}}
\def\ee{\end{equation}}
\def\bea{\begin{eqnarray}}
\def\eea{\end{eqnarray}}

\newcommand{\Photo}{}

\begin{document}
\vspace*{4cm}

\title{Observing dark matter clumps and asteroid-mass primordial black holes\\ in the solar system with gravimeters and GNSS networks}

%
%
%
%
%
%

\author{Bruno Bertrand$^1$, Michal Cuadrat-Grzybowski$^2$, Pascale Defraigne$^1$, Michel Van Camp$^1$ and S\'ebastien Clesse$^3$}
\address{$^1$Royal Observatory of Belgium, 1180 Brussels, Belgium\\ $^2$Space Engineering Department, Delft University of Technology, 2629 HS Delft, (the) Netherlands\\ $^3$Service de Physique Th\'eorique, Universit\'e Libre de Bruxelles (ULB), 1050 Brussels, Belgium.}

\maketitle\abstracts{
In this proceedings, we study the possible gravitational impact of primordial black holes (PBHs) or dark matter (DM) clumps on GNSS satellite orbits and gravimeter measurements.
It provides a preliminary step to the future exhaustive statistical analysis over 28 years of gravimeter
and GNSS data to get constraints over the density of asteroid-mass PBH and DM clumps inside the solar system. Such constraints would be the first to be obtained by direct observation on a terrestrial scale.}

\section{Introduction}


Whereas there are multiple indirect probes of the existence of Dark Matter (DM) from galactic to cosmological scales, there is no observational evidence coming from smaller scales.
It is plausible that DM sub-galactic clusters fragments into smaller parts, referred here as DM clumps. There are various theoretical scenarios in which a significant fraction of DM is made of dark objects: primordial black holes (PBHs)\cite{Hawking:1971ei,Carr:1974nx}, dark quark nuggets or strangelets~\cite{Witten:1984rs}, dark blobs or other composite states~\cite{Wise:2014jva,Grabowska:2018lnd}, axion or scalar miniclusters~\cite{Hogan:1988,Enander:2017ogx}, axion~\cite{Seidel:1993zk,Braaten:2018nag} or boson stars~\cite{Colpi:1986ye,Eby:2015hsq}...
This work focuses on 
DM clumps with a mass between 10$^{10}$ to 10$^{20}$ kg. This mass range is relevant because microlensing of stars becomes ineffective in detecting compact DM objects below 10$^{-19}$ kg~\cite{Montero-Camacho:2019jte}. 
For what concern the PBHs, a suspected mass threshold of $10^{11}$ kg comes from their evaporation through Hawking radiation~\cite{Auffinger:2022dic}. As those DM clumps travel in our galaxy, they eventually pass through the Solar System (SS), itself in motion around the Galactic center. 
In this work, we propose for the first time to jointly exploit gravimetry and global navigation satellite systems (GNSS) data in order to track anomalies in the gravitational potential induced by DM clumps passing sufficiently near the Earth. For GNSS, this is the first study addressing their direct gravitational influence on the satellite orbits. For gravimeters, we investigate the case of transient signatures whereas previous work only focused on the periodic signatures~\cite{Horowitz2019}. 

\section{DM clump and PBH in the solar system}

\begin{figure}[h]
    \centering
    \includegraphics[width=0.5\textwidth]{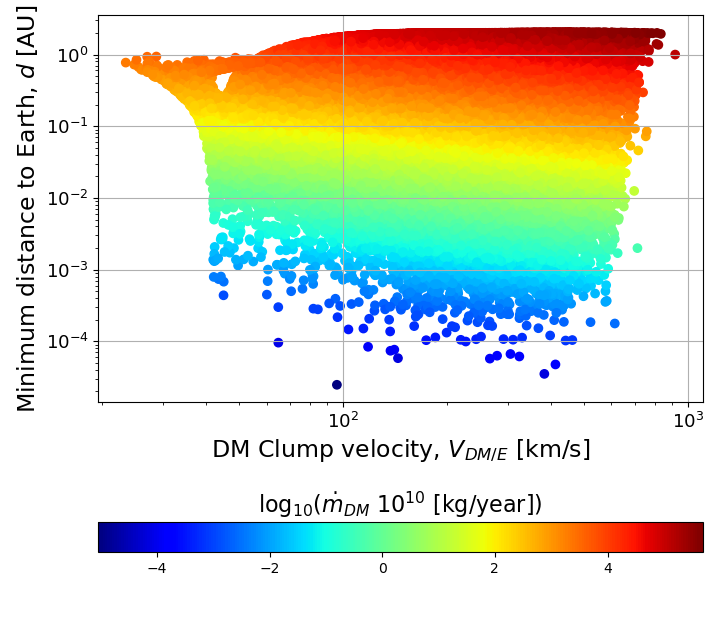}
    \vspace{-0.70cm}
    \caption{Monte-Carlo simulation of minimum Earth-Clump distance $d$ and velocity $V_{DM/ \Earth}$, with mass flow $\dot{m}_\textrm{DM}$ heat-map, for the range of impact parameters ($B \sim \mathcal{U}[0.98, 1.025]$ AU).}
    \label{fig:m_dot_year}
\end{figure}

Our modeling of events rate is based on the usual value of the DM density in the SS neighbourhood, $\rho_{\textrm{DM}} \simeq  0.4\, \textrm{GeV/cm}^3$. However, our analysis goes beyond simple approximations of a constant DM flux, by performing Monte-Carlo simulations of a realistic population of DM clumps with asteroid masses and by numerically integrating their trajectories in the SS. 
%
Their orbital motion is modeled as Keplerian, using a semi-analytical approach. 
%
Only two input parameters are sufficient to fully define the 2D orbit: the impact parameter $B$ and the hyperbolic excess velocity $V_{\infty}$.
Finally, the event rate and the phase space density close to the Earth are determined from the output parameters which are: the fly-by distance $d$, the relative velocity $V_{DM / \Earth}$ at the distance $d$, and the clump mass ${m}_\textrm{DM}$.
%
%
%
Fig.~\ref{fig:m_dot_year} shows the output of our MC simulation in terms of $d$ and the distribution of $V_{DM/ \Earth}$. The associated mass flux $\dot{m}_\textrm{DM}$ is presented as a heat-map.
Given that a model of point mass has been introduced, the notion of mass flux is independent of an underlying model for DM halo fragmentation. Indeed for example, a mass flux of $10^{10}$ kg/year cannot be distinguished from a single clump with a mass of $10^{12}$ kg every century at the same flyby distance $d$.
    \vspace{-0.65cm}
%


\section{Signature on GNSS and superconducting Gravimeters}

\subsection{GNSS}
Our modeling shows that any change in the gravitational potential induced by a DM clump should affect satellite orbits. 
%
We carried out a numerical computation of the gravitational signal caused by the DM clump on GNSS satellites, using the acceleration $\delta \vec{a}_\textrm{c}(t)$ in a third-body perturbation. 
This simulation is performed by propagating in time a 3D Keplerian osculating orbit for the GNSS satellite and one 3D unperturbed hyperbolic orbit for the DM object. Given these orbital paths, the orbit perturbation on GNSS satellite reads:
\begin{equation}\label{eq:da_dg_GNSS}
\frac{\delta \vec{a}_\textrm{c}(t)}{g_\oplus (t)} = - \frac{\mu_{\textrm{DM}}}{g_\oplus (t)} \left(\frac{\vec{r}-\vec{r}_\textrm{DM}}{|| \vec{r}-\vec{r}_\textrm{DM}||^3} + \frac{\vec{r}_\textrm{DM}}{||\vec{r}_\textrm{DM}||^3}\right)\, ,
\end{equation}
where $\vec{r}$ and $\vec{r}_\textrm{DM}$ are the positions of the GNSS satellite and DM Clump respectively, $\mu_{\textrm{DM}}$ is the DM celestial parameter and $g_\oplus (t)$ is the reference gravity field. The subsequent deviation of the GNSS reference orbit is performed by updating the Keplerian elements using Gauss' variational equations\cite{Aslanov2017}. The semi-major axis $a$ is chosen as observable since it is the most sensitive parameter:
\begin{equation}
\frac{\textrm{d}a}{\textrm{d}t} = \frac{2\, a^2}{h}\cdot \left[ e\, \text{sin}(\theta)\, \tilde{a}_r + \frac{h^2\, \tilde{a}_{\theta}}{\mu_\oplus\, r_0} \right]\, ,
\end{equation}
where $\mu_\oplus$ refers to the Earth's celestial parameter, $\vec{r}_0(t)$ is the initial reference orbit, $h$ is the orbital angular momentum, $e$ is the eccentricity, $\theta$ is the true anomaly and ($\tilde{a}_r$,$\tilde{a}_{\theta}$) are the radial and tangential components within the satellite frame of the satellite acceleration caused by the perturbation. 
%
\begin{figure}
    \includegraphics[width=.44\linewidth]{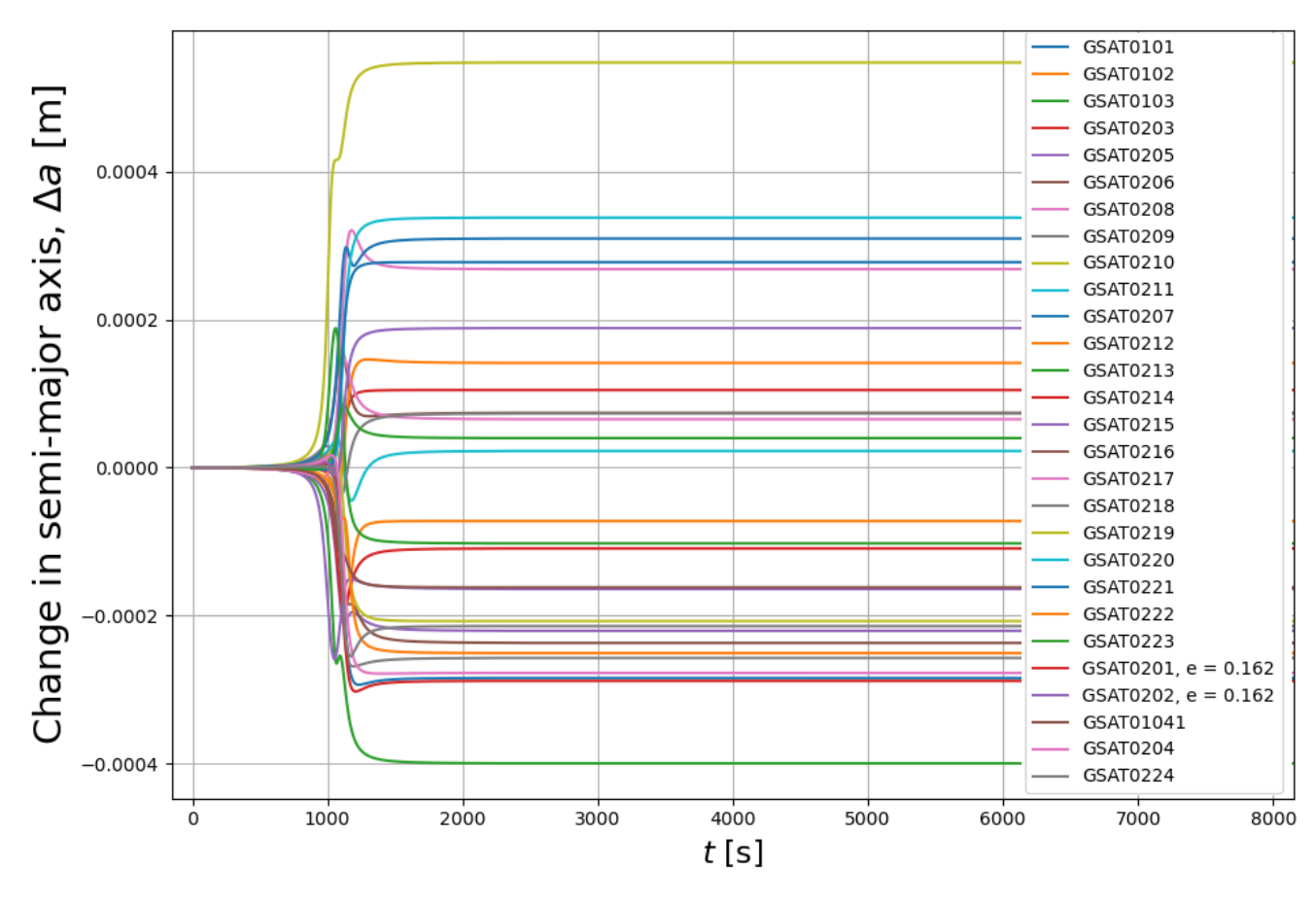}
    \includegraphics[width=.55\linewidth]{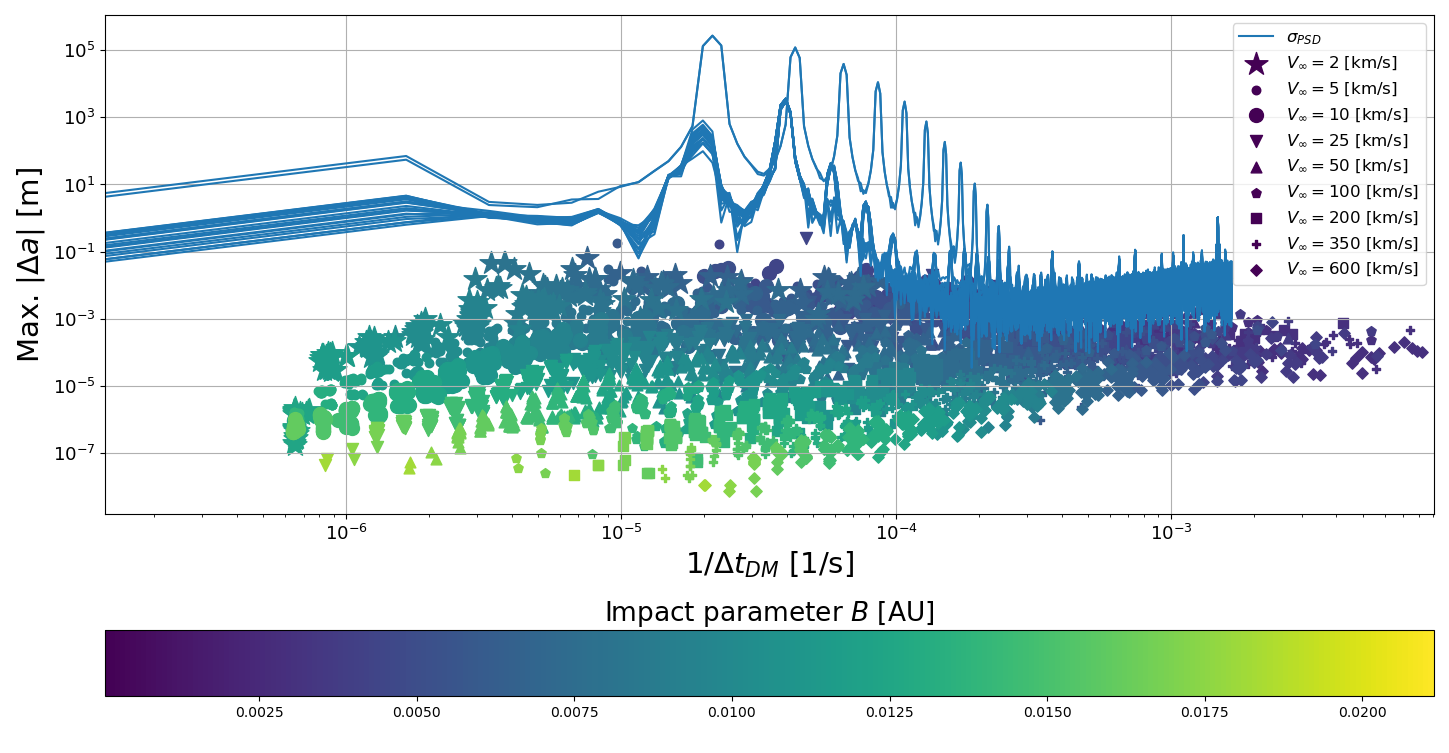}
    \vspace{-0.6cm}
    \caption{Signature of a DM clump of $10^{15}$ kg on the Galileo constellation. \textbf{Left:} Impulse response to the semi-major axis for a flyby at $B_\Earth=15500$ km, $V_{\infty} = 300$ km/s and $i_\textrm{DM} = 90$ deg. \textbf{Right:} Physical envelope of the max. $|\Delta a|$, realised for 9 excess velocities ranging from 2.5 km/s to 600 km/s) with each velocity associated to 20 values for $B_\Earth$ in the range [$10^{-4}$- $2\cdot 10^{-2}$] AU. Added to the latter is the constellation $\sigma_{PSD}$.}
    \label{fig:GNSS}
\end{figure}
The signature of a DM clump flyby on the Galileo Constellation is modeled in Fig.~\ref{fig:GNSS} (left), using $a$ as observable. The signal is a characteristic impulse, slightly dephased for each satellite, with a maximum near the Earth closest approach. 
Thanks to networks of fixed permanent GNSS stations, it is possible to determine the GNSS satellites orbit at the cm level. Such products are made available by e.g. the analysis center CODE\cite{Prange2020} for the International GNSS service (IGS). 
Fig.~\ref{fig:GNSS} (right) shows power spectral densities (PSD) of the semi-major axis of each GNSS satellite (solid lines), in combination with simulated events of duration $\Delta t_\textrm{DM}$ using DM clumps model orbits, with different values for $B_\Earth$ and $V_{\infty}$.

\subsection{Superconducting gravimeters (SG)}

\begin{figure}
    \includegraphics[width=.44\linewidth]{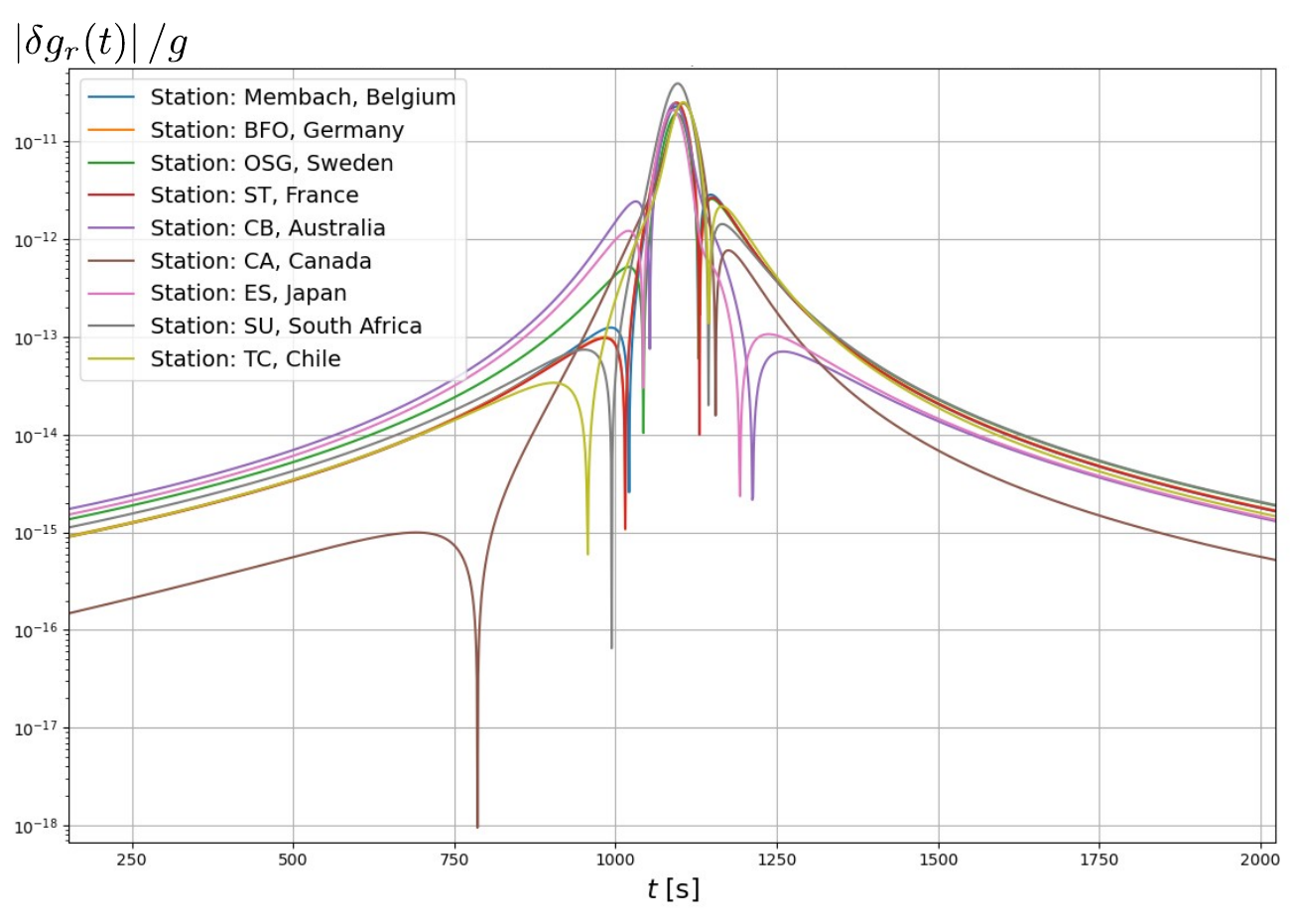}
    \includegraphics[width=.55\linewidth]{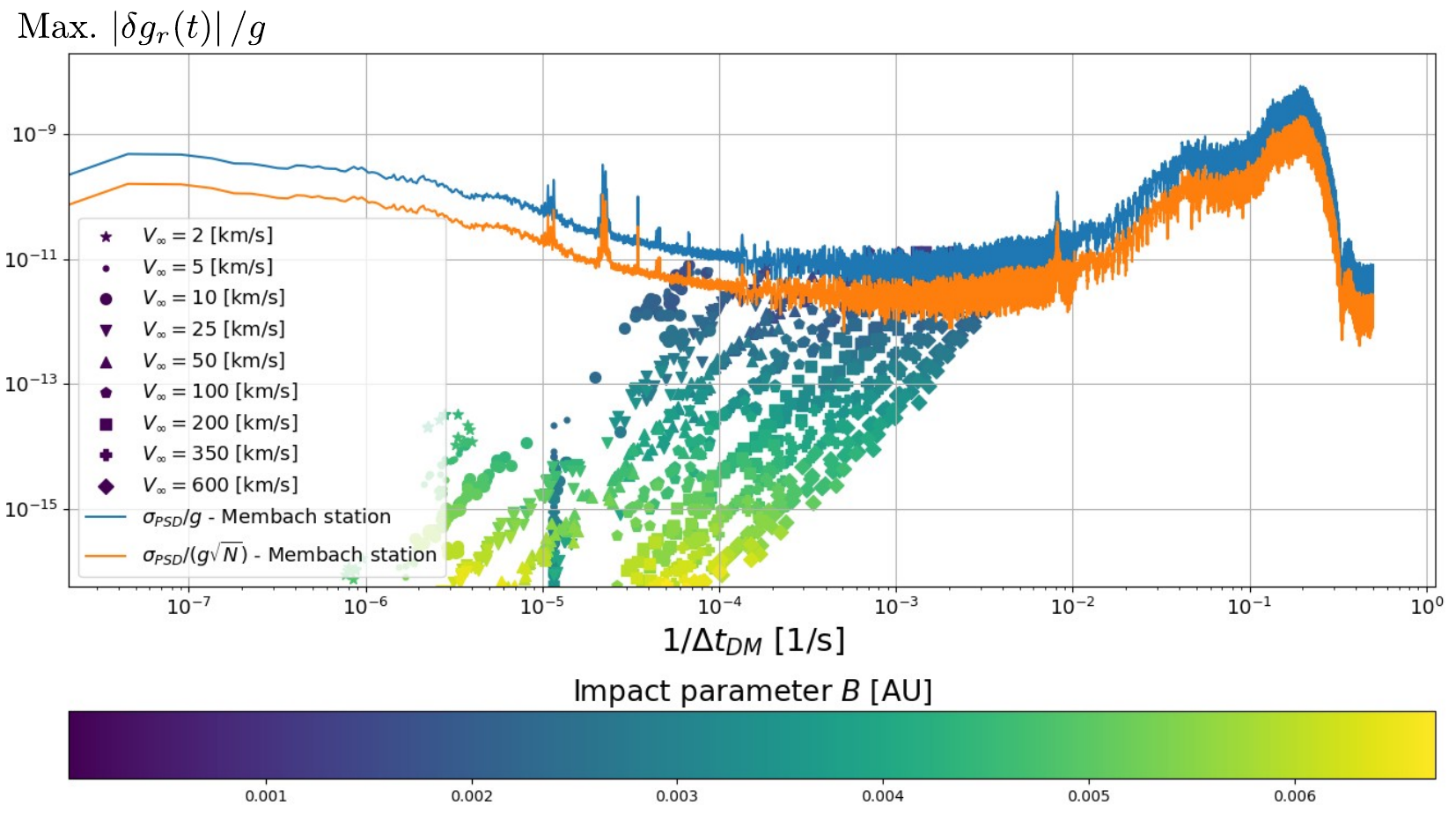}
    \vspace{-0.6cm}
    \caption{Signature of a DM clump of $10^{15}$ kg 
    \textbf{Left:} Gravitational perturbation induced by a flyby at $B_\Earth=15500$ km, $V_{\infty} = 300$ km/s and $i_\textrm{DM} = 0$ deg. 
    \textbf{Right:} Physical envelope of the max. $|\delta g_r/g|$, realised for 9 excess velocities ranging from 2.5 km/s to 600 km/s associated to 20 values for $B_\Earth$ in the range [$10^{-4}$- $10^{-2}$] AU. Added to the latter is the Membach station $\sigma_{PSD}/g$ (blue) with an additional $1/\sqrt{N}$ correction factor (Orange), where $N$ is the number of gravimeters.}
    \label{fig:gravimeter}
\end{figure}

For SG, the perturbation induced by a transient DM clump is measured as the radial component of the third-body perturbation. For flybys outside of the Earth, the relation for the third-body acceleration is similar to the one found for GNSS satellites (\ref{eq:da_dg_GNSS}).
Fly-bys inside of Earth occur when the Earth closest approach $d$ of the clump orbit is smaller than the Earth's radius. In that case, the time-dependent normalised gravimeter reading is computed as:
\begin{equation}\label{eq:Gravimeter}
\frac{\delta g_r(t)}{g} = \frac{1}{g} \,\left[ 
\mu_\textrm{DM}\,  \frac{\vec{r}_\textrm{g} - \vec{r}_{\textrm{DM}}}{||\vec{r}_\textrm{g} - \vec{r}_{\textrm{DM}}||^3} 
 + \frac{\textrm{d}^2X_{\Earth}}{\textrm{d}t^2} \frac{\vec{r}_{DM}}{||\vec{r}_{\textrm{DM}}||}\right] \cdot \frac{\vec{r}_\textrm{g}}{||\vec{r}_\textrm{g}||}\, , 
\end{equation}
where $\vec{r}_g$ is the position of the gravimeter and $\vec{r}_\textrm{DM}(t)$ is the modeled hyperbolic orbit. It is also assumed that $m_\textrm{DM}<<M_{enc}$ where $M_{enc}(r)$ is the Earth's enclosed mass at the radial distance $r_\textrm{DM}$. Two effects may be distinguished in (\ref{eq:Gravimeter}). The first one is a recoil of the Earth's center of mass, $d^2X_{\Earth}/dt^2$. The second effect is the actual acceleration caused by the DM clump.
The signature of a DM clump flyby on the gravimeter residuals is modeled in Fig.~\ref{fig:gravimeter} using a worldwide network of 9 gravimeters. The nature of the perturbation is a transient peak reached near the Earth closest approach. 
The level of precision is of the order of $10^{-11} g$ within 1 minute, where $g$ is the Earth gravitational acceleration. 
%
Fig.~\ref{fig:gravimeter} (right) shows the PSD of residuals (observation - tidal variations) time series in the Membach station~\cite{VanCamp:2005} in Belgium, in combination with simulated events of duration $\Delta t_\textrm{DM}$ using DM clumps orbit models, with different values for $B_\Earth$ and $V_{\infty}$.


\section{First results}

\begin{figure}
    \includegraphics[width=.40\linewidth]{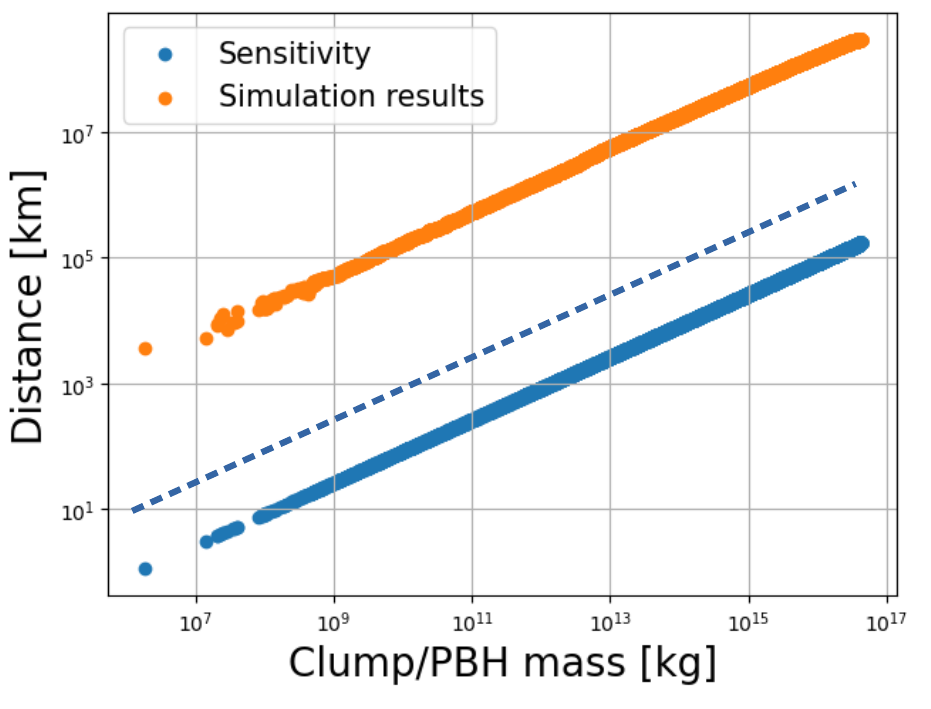}
       \centering
    \vspace{-0.1cm}
    \caption{Orange dots: Simulation result of DM clump distance and integrated mass computed over a $20$ year-period obtained from the Monte-Carlo simulation. 
    Blue dot: One-probe sensitivity distance as a function of the accumulated clump mass. Blue dashed line: Expected sensitivity after a full correlation analysis.}
    \label{fig:Sensitivity}
\end{figure}

A preliminary analysis based on the PSD of the gravimeter residuals and Galileo orbital solutions provides a first assessment of the `one-probe', i.e. one satellite or one gravimeter, sensitivity. This sensitivity is determined by the overlapping area between the PSD and the simulated events in Fig.~\ref{fig:GNSS} and~\ref{fig:gravimeter}.~ Focusing on GNSS, Fig.~\ref{fig:Sensitivity} relates the single-satellite sensitivity (blue dots) to a minimum clump mass and a corresponding flyby distance.
%
The ratio with our simulated event rates (orange dots, Fig.~\ref{fig:Sensitivity}) is around 6 orders of magnitude, so that a single probe is not enough to hope a detection. 
%
%
However, this single probe analysis paves the way for a future statistical analysis based on correlations between a constellation of probes using the 28-years of publicly available GNSS and gravimeter data. 
Such a statistical analysis would enable to gain between 2 and 3 orders of magnitude in sensitivity (dashed blue line, Fig.~\ref{fig:Sensitivity}).
%

Finally, our one-probe sensitivity implies a state of equilibrium with a total DM clump mass of less than 0.1 Ceres mass in a 1.5 AU sphere around the Sun. This one-probe results would not compete with the bounds based from space probes and planet ephemeris in the SS~\cite{Pitjev:2013sfa}. However, these latter limits apply only to a permanent cloud of particles in the SS and are not transferable to transient DM clumps. Anyway, a full statistical analysis of GNSS and SG data would surpass the limits based on SS ephemeris~\cite{Pitjev:2013sfa} by at least one order of magnitude. 
Hence, such constraints would be the first and the best to be obtained by direct observation on a terrestrial scale, before LISA is operational~\cite{Baum:2022duc}. 


\section*{References}

\bibliography{biblio.bib}

%
%
%
%

\end{document}